# Distributions of Peak Flux and Duration for Gamma-Ray Bursts

Vahé Petrosian[1] and Theodore T. Lee[2]

Center for Space Science and Astrophysics, Stanford University, Stanford, CA 94305-4055

## ABSTRACT

Many of the important conclusions about Gamma-Ray Bursts follow from the distributions of various quantities such as peak flux or duration. We show that for astrophysical transients such as bursts, multiple selection thresholds can lead to various forms of data truncation, which can strongly affect the distributions obtained from the data if not accounted for properly. Thus the data should be considered to form a multivariate distribution. We also caution that if the variables forming the multivariate distribution are not statistically independent of each other, further biases can result. A general method is described to properly account for these effects, and as a specific example we extract the distributions of flux and duration from the BATSE 3B Gamma-Ray Burst data. It is shown that properly accounting for the aforementioned biases tends to increase the slope of the $\log N$-$\log S$ relation at low values of $S$, and dramatically increases the number of short duration bursts.

*Subject headings:* gamma rays: bursts

## 1. Introduction

The unusual nature of Gamma-Ray Bursts (GRBs) has provided many challenges for astronomers in the design of instruments and observations for their detection, in analysis of the data, and in theoretical modeling. The primary sources of these difficulties are their extreme variability, short durations ($< 10^3$ s), varied "light curves," and the apparent absence of any corresponding lower energy photons. Consequently there exist no simultaneous observations at other wavelengths, especially those accessible to ground based observatories which could allow identification of counterparts and provide knowledge of their distances, luminosities, and spatial distributions. As is the case in all other astronomical situations, the absence of direct knowledge of distance means that we must rely on the so-called $\log N$-$\log S$ distribution to infer the combined spatial and luminosity distribution $\Phi(L, r)$.

---

[1] Departments of Physics and Applied Physics

[2] Department of Applied Physics

For steady sources, with $S$ standing for the flux $f = L/4\pi r^2$, the number of sources at flux $f$ is related to the luminosity function $\Phi$ as

$$n(f) = \int_0^\infty \Phi(L,r) \frac{dV}{dr} 4\pi r^2 dr, \qquad (1)$$

where $V(r)$ is the volume (co-moving for cosmological sources) of space occupied up to distance $r$ (the luminosity distance for cosmological sources). From comparisons of $n(f)$ with observations one hopes to determine the luminosity function and the geometry of the space occupied by the sources. This of course cannot be done uniquely from the $\log N$-$\log S$ analysis alone, and we must rely on other information for a complete picture. Nevertheless it is clear that the $\log N$-$\log S$ distribution is an important tool for the study of GRBs, and it is imperative that it be determined accurately from observations.

For GRBs, and variable sources in general, care is required for an accurate and bias-free determination of their distributions. When a source can vary over a time scale shorter than the length of accumulation of data, there is the uncertainty whether to use peak flux or total (time integrated) flux (i.e. fluence) for the parameter $S$. GRBs, with durations ranging from $10^{-2}$ to $10^3$ s, clearly fall into this category and the earlier analyses of their $\log N$-$\log S$ relation suffered from this uncertainty. This difficulty was somewhat alleviated by analyses of the data in terms of $c = \bar{C}_P/\bar{C}_{lim}$ (or equivalently in terms of the so-called $V/V_{max} = c^{-3/2}$), where $\bar{C}_P$ is the peak count rate (averaged over a predetermined trigger interval $\Delta t$) and $\bar{C}_{lim}$ is the variable threshold rate (which also depends on $\Delta t$). Unfortunately, as pointed out by Petrosian (1993; hereafter PI) the distribution of the ratio $c$, aside from being capable of testing the simplest model (a homogeneous, isotropic distribution of sources in a static, Euclidean geometry, HISE for short), is of little use in the determination of $\Phi$ or $V(r)$ in equation (1). As shown in PI, the distribution of $c$ is related in a more complicated way to $\Phi$ and $V$, involving a convolution of equation (1) with the distribution of $\bar{C}_{lim}$. It was also shown in PI that such a complication is unnecessary; the distribution of $\bar{C}_P$ (or the peak flux $\bar{f}_P$) can be obtained directly from the observed bivariate distributions of $\bar{C}_P$ and $\bar{C}_{lim}$, with correct accounting of the bias introduced due to variable $\bar{C}_{lim}$.

The variability in the threshold rate and the finite triggering interval $\Delta t$ introduce additional complications and selection effects which need to be considered. Many instruments, including BATSE on the *Compton Gamma-Ray Observatory*, define a burst to

be occurring when the average photon count rate

$$\bar{C}(t) = \int_t^{t+\Delta t} C(t)dt/\Delta t \qquad (2)$$

exceeds the threshold rate $\bar{C}_{lim}$. Bursts are characterized by $\bar{C}_{lim}$ and the maximum of this rate $\bar{C}_P = \max[\bar{C}(t)]$. BATSE uses three trigger intervals $\Delta t = 64$, 256 and 1024 ms, with corresponding values of $\bar{C}_{lim}$ determined for each value of $\Delta t$. The bias introduced by this process, even if $\bar{C}_{lim}$ were constant, occurs because bursts could have true durations $T$ larger or shorter than $\Delta t$. As pointed out by Petrosian, Lee & Azzam (1994; hereafter PII), for $T \gg \Delta t$ the average $\bar{C}_P$ is a good representation of the true peak rate $C_P = \max[C(t)]$, but for bursts with $T \ll \Delta t$ the average rate $\bar{C}_P$ is proportional to the total count fluence $\mathcal{F}_C = \int_T C dt$ as $\bar{C}_P = \mathcal{F}_C/\Delta t$ and clearly is an underestimation of the true $C_P$, with the degree of underestimation increasing with decreasing $T/\Delta t$. Therefore the observed distribution must be characterized by at least three variables: $\bar{C}_P$, $\bar{C}_{lim}$, and $T$. Furthermore, it is also clear that the above bias depends on the shape of the light curve. For example, bursts with $T \gg \Delta t$ which have a spiky peak with a time scale shorter than $\Delta t$ also suffer from the above bias.

Thus in reality the task at hand is more complicated. We need to extract the univariate distributions $n(C_P)$, $g(T)$, etc. from a multivariate observed distribution $\psi(\bar{C}_P, \bar{C}_{lim}, T, \alpha_i, \ldots)$, where $\alpha_i$ are parameters characterizing the shape of $C(t)$. Care is necessary in such an extraction because the data is truncated, sometimes simply but sometimes in a complicated way, by the triggering and other observational selection effects, and because the variables of these multivariate distributions may not be independent of each other (they might be correlated).

In §2 we describe the difficulties that arise because of these effects and describe methods that we have developed to overcome them and obtain accurate distributions. Applications of these methods to the data from the BATSE 3B catalog are described in §3, where we derive the distributions of peak flux and duration. As already was evident from our preliminary application of these methods to the 1B catalog in PII, we find that ignoring these complications gives misleading results. Finally in §4 we present a summary and discussion of our results.

Note that usually this analysis is done in terms of the cumulative distribution

$$N(f) = \int_f^\infty n(f')df' = \int_0^\infty \frac{dV}{dr}dr \int_{4\pi r^2 f}^\infty \Phi(L,r)dL. \qquad (3)$$

The differential distribution with its independent points is preferred. We treat the problem in terms of the differential distribution or the logarithmic slope

$$s(f) = -\frac{d\log N}{d\log f} = \frac{fn(f)}{N(f)},\tag{4}$$

which in spite of its appearance represents a set of independent slopes. Further details on this aspect of the problem have been presented by Efron & Petrosian (1992) and Azzam & Petrosian (1996).

## 2. Obtaining Univariate Distributions From Multivariate Data

The problems mentioned above associated with analysis of GRB data, and in many other astronomical data analyses, can be reduced to the generic case of the determination of a bias-free univariate distribution from a multivariate distribution function. To simplify the discussion in what follows in this section we will consider the case where the multivariate distribution has only two variables and can be written as $\psi(x, y)$. The generalization to the multivariate situation only adds computational complexity with little conceptual difficulty. Furthermore, let us assume that $\psi$ is separable, such that

$$\psi(x, y) = f(x)g(y).\tag{5}$$

In order to obtain univariate distributions which are unbiased, two major effects must be taken into account. The first is the effect of data truncation due to observational selection processes. The second is the correlation that could be present between the variables ($x$ and $y$ in this presentation).

### 2.1. Effects of Truncation

Because of the limited sensitivity of experimental or observational procedures, all data are essentially truncated by the sensitivity thresholds. Namely, the available data is limited to say $x \geq x_{lim}$ and $y \geq y_{lim}$ (the situation with upper limits can be treated similarly and will not be considered here). The simplest truncations, in which $x_{lim}$ and $y_{lim}$ are constants, pose no additional problem. Data with only these kinds of truncations parallel to the axes are sometimes referred to as untruncated. Difficulty arises when one (or both) of these thresholds is a function of the other variable. For example, if threshold $x_{lim}$ is constant but the detection threshold of $y$ is a function of $x$, say $y_{lim} = h(x)$, then the data are called *truncated* and the distributions of observed $x_i$ and $y_i$ (where $i = 1, 2, \ldots n$, and $n$ is the total number of data points) are biased and do not represent the true distributions

$f(x) = \int_0^\infty \psi dy$ or $g(y) = \int_0^\infty \psi dx$. Note that this general truncation can be reduced to the generic case by transformation of the variables $x' \to h(x), y' \to y$ resulting in the generic truncation $y' > x'$.

The GRB data suffers from at least two truncations. The well known one is the requirement that $\bar{C}_P > \bar{C}_{lim}$, arising from the variable threshold $\bar{C}_{lim}$. The usual practice to account for this truncation has been to treat the data in terms of $c = \bar{C}_P/\bar{C}_{lim}$ or $V/V_{max} = c^{-3/2}$. This is equivalent to the transformations $x'' \to x$ and $y'' \to y/h(x)$ in the above example, which reduces to the case of simple truncation parallel to the axes. As mentioned in §1, this treatment has a limited utility and is shown in PI to be unnecessary because one can account for this bias and obtain the distributions of $\bar{C}_P$ and $\bar{C}_{lim}$ directly. Another bias, the subject of PII, is due to the finite duration $\Delta t$ of the burst trigger process. This gives rise to a peak photon count rate threshold which is a function of duration, i.e. to a truncation in the $\bar{C}_P$–$T$ plane. As also shown in PII, the effect of this truncation can also be accounted for by using the same methods used for the truncation due to the variation of $\bar{C}_{lim}$. We will return to these biases in the next section.

### 2.2. Effects of Correlations

The second major effect, which has not been recognized or acknowledged widely, is that even when the truncation is simple, i.e. parallel to the axes ($x_{lim}$ and $y_{lim}$ independent of $y$ and $x$, respectively), the observed distributions of $x_i$ and $y_i$ will be biased if $x$ and $y$ are correlated. For example, suppose that the average value of $y$ varies with $x$ as $\langle y \rangle = k(x)$, so that we can write

$$\psi(x,y) = f(x)g(y/k(x))/k(x). \tag{6}$$

It is then easy to show that for $y > y_{lim}$ the observed distribution of $x$ is

$$f_{obs}(x) = \int_{y_{lim}}^\infty \psi(x,y)dy = f(x)G(y_{lim}/k(x)), \tag{7}$$

where

$$G(u) = \int_u^\infty g(u')du', \quad G(0) = 1, \tag{8}$$

which is in general different from the true distribution $f(x)$. The difference between $f(x)$ and $f_{obs}(x)$ decreases with decreasing dispersion in the values of $x$ and $y$. Only in the case of perfect correlation with no dispersion ($g(u) = \delta(u)$) will $f_{obs}(x)$ be proportional to $f(x)$. It is clear that a similar discrepancy will be present between the true and observed distributions of $y$.

In the case of GRBs, if $\bar{C}_P$ and $\bar{C}_{lim}$ (or $\bar{C}_P$ and $T$) are correlated we cannot ignore the distributions of $\bar{C}_{lim}$ or $T$ in the determination of $n(\bar{C}_P)$, which has been a ubiquitous practice in previous studies. Note that such correlations cannot be ignored even when dealing with the distribution of the ratio $c = \bar{C}_P/\bar{C}_{lim}$. Hence, the well-known deviation of the logarithmic slope of the cumulative distribution of $c$ from -3/2 or the deviation of $\langle V/V_{max} \rangle$ from 0.5 does not necessarily indicate a deviation from HISE unless it is known that $\bar{C}_P$ and $\bar{C}_{lim}$ are uncorrelated with each other or $T$. All of the far-reaching consequences derived from the analysis of the distribution of $c$ or $V/V_{max}$ are based on this seemingly untested assumption.

Clearly both the effects of truncation and correlation must be taken into account when dealing with multivariate data that is *truncated* and its variables are *correlated*. When truncation is included in the above analysis, equation (7) becomes, upon replacement of $y_{lim}$ with $h(x)$,

$$f_{obs}(x) = f(x)G(h(x)/k(x)). \tag{9}$$

This replacement further differentiates between the observed and true distributions, unless the correlation form $k(x)$ is known *a priori* and the truncation is chosen parallel to it, namely $h(x) \propto k(x)$. It should also be noted that if the degree of correlation between the variables and the distribution of one of the variables is known the true distribution of the other variable can be determined from the observations and equation (9). Unfortunately such *a priori* knowledge often does not exist, and generally this information must be deduced from the observed data itself.

### 2.3. The Basic Non-Parametric Method

There are many ways of obtaining univariate distributions from bivariate or multivariate data. Most rely on parametric fits to binned data. There are two problems with these methods, beyond those mentioned above in regards to truncation and correlation. First, *a priori* parameterization is arbitrary and can give misleading results. Second, binning amounts to smoothing the data, which often means that all the data cannot be used and results in loss of information on scales smaller than those used for the binning. Of course, in the presentation of results or when comparing with theoretical models it may be necessary and convenient to bin the results. However, this should be done only after all the data is utilized to its maximum extent. Therefore, a non-parametric analysis of the data and the avoidance of binning (or the choice of bin sizes which are smaller than the observational errors) are preferred. The method that we have developed and applied to GRB data in the past and for this paper is non-parametric and does not use any binning.

As shown in the previous two subsections, a correct procedure for the determination of the distributions requires first and foremost an accurate measure of the degree of correlation between the variables in a truncated data set. Standard correlation tests fail in the presence of truncations, which may introduce artificial correlations. In our method, both the determination of correlations and the extraction of the univariate distributions once the correlation is known rely on the concept of the *associated set*.

Consider a set of data points $(x_i, y_i)$, $i = 1, 2 \ldots n$, ordered in $y$ such that $y_i > y_{i+1}$, which is truncated according to the relation $y_i \geq h(x_i)$. For each data point, say $y_i$, there exists an *associated set* of points $\mathcal{M}_i$ which is defined to consist of all data points within the largest untruncated (i.e. simply truncated) region. These are points $(x_j, y_j)$ with $y_j > y_i$ and $x_j$ such that $h(x_j) < y_i$, and lie in a box with sides parallel to the $x$ and $y$ axis (sometimes these sets are defined with $y_j \geq y_i$ so that they include the data point $y_i$ in question). Clearly, we can define similar sets, say $\mathcal{N}_i$ associated with points $x_i$, where $\mathcal{N}_i = \{(x_j, y_j) : x_j < x_i \text{ and } y_j > h(x_i)\}$. Boxes containing sample associated sets $\mathcal{M}_i$ and $\mathcal{N}_i$ are shown in Figure 1 for the bivariate distribution of peak and threshold fluxes $\bar{f}_P$ and $\bar{f}_{lim}$ as defined below for the BATSE 3B 1024 ms catalog.

If the variables $x$ and $y$ are uncorrelated or are stochastically independent, then the various characteristics of the distributions can be expressed as functions of the number of points in the associated sets. In fact, for any given point $y_i$, the reciprocal of the number of points $M_i$ in the associated set $\mathcal{M}_i$ is a direct measure of the logarithmic slope of the cumulative distribution $G(y)$ at $y_i$. It can be shown (for details see Efron & Petrosian 1992 and Azzam & Petrosian 1995) that the logarithmic slope at $y_i$ is given by

$$s_{G,i} = \frac{y}{G}\frac{dG}{dy} = \frac{\delta(y - y_i)}{M_i + \Theta(y - y_i)}, \qquad (10)$$

where $\delta$ is the Dirac delta function, and $\Theta$ is the Heaviside step function. It follows that the cumulative distribution $G$ is obtained from the data as

$$G(y_i) = G(y_1) \prod_{j=2}^{i} (1 + M_j^{-1}), \quad i > 1, \qquad (11)$$

which is the relation obtained from the method first proposed by Lynden-Bell (1971) and later elaborated on by Woodroofe (1985) and Petrosian (1986). $G(y_1)$ is the value of $G$ for the first point, which is unknown since by definition there are no data points with $y > y_1$. It turns out that all non-parametric methods in the limit of one object per bin reduce to this simple method (for details see the review by Petrosian 1992). From equations (4), (10), and (11) we can obtain the differential distribution $g(y) = G(y_i)s_{G,i}/y_i$. Note that independently of these distributions we can obtain the slope $s_{F,i}$, the cumulative distribution $F(x_i) = \int_0^{x_i} f(x)dx$, and $f(x_i)$ at all values of $x_i$ in terms of the unknown value of $F(x_1)$ at the lowest observed point. Whether the points are arranged in increasing or decreasing order is a matter of convenience or preference.

The concept of the associated set is also used in the test for correlation. Since the truncation of the data can produce an artificial correlation between the observed values of $x_i$ and $y_i$, a simple correlation test will lead to misleading results. Only correlations involving untruncated sets of points, i.e. *associated sets*, are real. Efron & Petrosian (1992) describe a test of independence based on rank ordering in the associated sets. One outcome of such a test would be a parametric determination of correlation, e.g. the function $k(x)$ introduced above (see Lee, Petrosian, & McTiernan 1993 & 1995 for examples of this). Once $k(x)$ is known, the transformation $y' \to y/k(x)$ would lead to a new set of uncorrelated variables $(x, y')$, the distributions of which can be obtained from equations (4), (10), and (11).

## 3. BATSE Flux-Duration Distributions

The importance of the distribution of peak fluxes (photon count rates) has already been stressed in the introduction. The distribution of burst durations also gives important information about the nature of the sources. For example, analysis by the BATSE team (Kouvelieutou et al. 1993) of the raw distribution of durations for bursts in the 1B catalog (without regard for the selection effects and potential biases due to correlations) indicates the possible existence of two populations of bursts, with the separation between these occurring at a burst duration $T \approx 2$ s. This separation is less clearly evident in the 3B catalog raw data. It is therefore important to determine how the effects we are considering here change this picture. Furthermore, the determination of the degree of correlation between the peak fluxes and durations of the GRBs is important not only for finding the individual distributions of these two characteristics but also for the so-called cosmological time dilation test. If the absolute peak luminosity of GRBs is constant (a very unlikely hypothesis considering the large dispersion in the pulse shapes and durations of GRBs), then a cosmological distribution of sources should show an anticorrelation between duration and peak flux. Weaker sources, being at higher redshifts, would be time dilated by larger amounts. Norris et al. (1994 & 1995) show that there exists such a relationship for a subset of GRBs observed by BATSE. These studies examined only a small fraction of the total number of bursts, for which the truncation effects were unimportant. The reality of this result and its interpretation as simple cosmological time dilation has been questioned (Band 1994; Mitrofanov 1994; Fenimore et al. 1995). Using our methods, we may examine this relation for all GRBs for which fluxes, flux limits, and durations are known. This data set comprises $\sim 50\%$ of all bursts, as compared to the $\sim 20\%$ used in the Norris et al. studies.

As shown below and described in greater detail in another publication (Lee & Petrosian 1996a), we cannot confirm the Norris et al. results with the publicly available BATSE catalog data.

Before we proceed further we must define what we mean by the duration of a burst. It is well recognized that GRBs do not have well-defined pulse profiles. Light curves are as varied as the number of the bursts. In the distribution $\psi(\bar{C}_P, \bar{C}_{lim}, T, \alpha_i)$ we would require many parameters $\alpha_i$ to describe all of the observed pulse shapes. Therefore, it is futile to attempt to solve this problem exactly. We are forced to average over all possible light curves and reduce the distribution to a trivariate one, $\psi(\bar{C}_P, \bar{C}_{lim}, T)$. There are various ways one can define a pulse duration. The BATSE catalogs provide two duration estimates, $T_{90}$ and $T_{50}$, which are the intervals of time containing 90% and 50% of the burst fluence.

For simple pulses these and other definitions of duration are related to and are often proportional to each other and to the total duration or characteristic time scale of the pulse. For example, for a square pulse $T_x = T(x/100)$ and for a triangular (one or two sided) pulse $T_x = T(1 - \sqrt{1 - x/100})$, where $T$ is the total duration. Similarly, for an exponential pulse of time constant $\tau$ we have $T_x = \tau \ln(100/(100 - x))$. For more complicated pulses consisting of several overlapping or well separated spikes, $T_{90}$ could be a measure of the separation between spikes, while $T_{50}$ could be a measure of the duration of the main pulse of the burst.

### 3.1. Data Truncation Effects

The observed distributions of $\bar{C}_P, \bar{C}_{lim}$, and $T$ are subject to two truncations. The first arises because $\bar{C}_P \geq \bar{C}_{lim}$, where $\bar{C}_{lim}$ is variable. The second, the existence of which and its effect were pointed out in PII, arises because of the finite duration of the trigger interval $\Delta t$. BATSE uses three values, 64, 256 and 1024 ms, for $\Delta t$. A burst occurs if the total instrument counts accumulated during one of these intervals exceeds the predetermined value of the threshold counts $\bar{C}_{lim,\Delta t}$. The peak counts listed in the catalog are the largest values of

$$F(t, \Delta t) = \int_t^{t+\Delta t} C(t')dt' \qquad (12)$$

for the triggered bursts. If we define an average count rate $\bar{C}_{\Delta t}(t) = F(t, \Delta t)/\Delta t$ (as in Eq. [2]), then the peak photon count rate $\bar{C}_{P,\Delta t}$ is the maximum of $\bar{C}_{\Delta t}(t)$ (from now on $\bar{C}_P$ should be understood to implicitly mean $\bar{C}_{P,\Delta t}$). This however is not necessarily equal to the true peak rate $C_P$, which is the maximum of the light curve $C(t)$.

There are two effects which make $\bar{C}_P \neq C_P$. The first and the more important bias causes $\bar{C}_P < C_P$ for any burst which varies significantly over time scales less than $\Delta t$ (namely short duration bursts or bursts with a dominant short duration spike). In particular, for bursts with total duration $(T < \Delta t)$, $\bar{C}_P \Delta t$ is equal to the count *fluence* $\mathcal{F}_C = \int_T C(t) dt$ of the burst. On the other hand, for bursts with durations (or time scales of the strongest spike) $T \gg \Delta t$, the average $\bar{C}_P$ is a good measure of the true peak rate $C_P$. For a simple pulse shape it is easy to see that $\bar{C}_P/C_P \leq 1$, with the ratio decreasing rapidly for pulses with durations less than $\Delta t$. For example, for square, triangular and exponential pulses with duration or time constant $T$ we have, respectively,

$$\bar{C}_P = \begin{cases} C_P & \text{for } T \geq \Delta t \\ C_P T/\Delta t = \mathcal{F}_C/\Delta t & \text{for } T \leq \Delta t \end{cases} \tag{13}$$

$$\bar{C}_P = \begin{cases} C_P(1 - \Delta t/2T) & \text{for } T \geq \Delta t \\ C_P T/2\Delta t = \mathcal{F}_C/\Delta t & \text{for } T \leq \Delta t \end{cases} \tag{14}$$

$$\begin{aligned} \bar{C}_P &= C_P(1 - e^{-\Delta t/\tau})/(\Delta t/\tau) \\ &= \begin{cases} C_P(1 - \Delta t/2\tau) & \text{for } \tau \gg \Delta t \\ C_P \tau/\Delta t = \mathcal{F}_C/\Delta t & \text{for } \tau \ll \Delta t \end{cases} \end{aligned} \tag{15}$$

The above expressions could be off slightly due to bin edge effects, with the maximum error a factor of two in the unfortunate case where exactly half of the fluence of a short $(T < \Delta t)$ burst happens to be located exactly on the edge between two integration periods of time $\Delta t$. Also note that the second expression in each case shows that $\bar{C}_P \Delta t = \mathcal{F}_C$, the count fluence, and is not a measure of peak photon count rate. Thus, $\bar{C}_P$ as determined by BATSE is a measure of peak rate only for long duration and gradual events (those with time scales $T > \Delta t$), while it is a measure of the fluence for short duration bursts. For bursts with long durations but containing a dominant spike with duration $T_{spike} < \Delta t$, $\bar{C}_P$ is an underestimation of $C_P$. Such bursts could greatly complicate the issue. However, as we shall see below, the majority of bursts do not have this troublesome characteristic. Therefore, in any analysis of this data we should treat the short and long duration (relative to trigger time $\Delta t$) bursts separately or find a method to correct for this bias. From examination of equations (13), (14), and (15) we can see that a rough approximation to correcting for the bias would be to make the transformation $\bar{C}_P \to C_P$, with

$$\frac{C_P}{\bar{C}_P} = \left(\frac{T + \Delta t}{T}\right), \tag{16}$$

which to within a factor of less than 2 agrees with all three pulse shapes. $C_P$ would then be our best estimate of the true peak count rate.

We empirically test the above relation with the BATSE data. Since two of the BATSE trigger times $\Delta t$ (64 ms and 1024 ms) are widely separated, for bursts with $T$ or $\tau > 64$ ms, $\bar{C}_{P,64}$ will be a good measure of the true $C_P$ so that the ratio $\bar{C}_{P,64}/\bar{C}_{P,1024}$ would provide a good estimate for the right hand side of equation (16) for $\Delta t = 1024$ ms. For very short bursts ($T_{90} < 64$ ms), both the 64 ms and 1024 ms triggers give us a measure of the fluence $\mathcal{F}_C$ so that the ratio $\bar{C}_{P,64}/\bar{C}_{P,1024} \to 16$ and is no longer a measure of the right hand side of the above equation. The top panel of Figure 2 shows this ratio (which was called the *variability* by Lamb, Graziani, & Smith 1993) for the BATSE 3B data as a function of $T_{90}$.

As expected for $T_{90} > 1024$ ms and $T_{90} < 64$ ms this ratio tends towards the asymptotic values of 1 and 16 (though because of the very same bias there are very few bursts with $T < 64$ ms). For $64 < T_{90} < 1024$ ms the general trend of the data is very similar to the expression on the right hand side of equation (16), which is depicted by the solid line. The exact expression for the three simple pulse shapes (using the correct relation between $T_{90}$ and $T$) are shown by the three dashed curves in Figure 2 (bottom panel).

Aside from the relatively good agreement it should be noted that there are very few bursts in the upper right hand corner of this figure. Bursts which have a long overall duration but contain a sharp spike (with time scale < 1024 ms) would occupy this part of the diagram. As mentioned above such bursts would greatly complicate the bias correction. The absence of such bursts indicates that the dominant spike in a long duration burst tends to be broader than 1024 ms ($C(t)/dC(t)/dt > 1024$ ms near the peaks). However, the deviation from unity of the ratio at the long duration end indicates that this is not always true. Undoubtedly some of the vertical dispersion is due to random fluctuations and observational errors. In addition, some of the deviation from unity could be due to a second bias in the data, which in contrast to the above is important for slowly varying (not spiky) bursts. This bias is discussed by Lamb et al. (1993), who attribute it to C. Meegan. This so-called peak flux bias is a minor effect and arises from statistical fluctuations which could potentially give a peak count rate $\bar{C}_P \geq C_P$ for bursts with small $C_P$ and long durations ($T \gg \Delta t$). In appendix A we discuss the peak flux bias in more detail and carry out some simulations to determine its effect. The bottom panel of Figure 2 shows the same data as in the top panel (dotted histogram), except that it has been median filtered in $T_{90}$ in groups of 3 bursts for the purpose of clarity. The solid histogram shows the $\bar{C}_{P,64}/\bar{C}_{P,1024}$ ratio "corrected" for the peak flux bias under the assumption of square pulses, which results in the maximum difference between the observed and true ratios. The theoretical values of the ratio for the pulse shapes given by equations (13)–(15) are also shown on this diagram. Given the uncertainties inherent in the pulse shapes of the bursts and the correction for the peak flux bias, we take equation (16) to be a good fit to the data and in what follows we use this to correct for the main bias discussed above.

## 3.2. Testing for Correlations

Having determined the various truncations of the trivariate data, we can now proceed with the determination of the distributions of individual variables. As stressed in §2, the first step in the analysis of multivariate data is testing the data for any correlation between the variables $\bar{C}_P$, $\bar{C}_{lim}$ and $T$. The BATSE catalog, in addition to providing $\bar{C}_P$ and $\bar{C}_{lim}$, also gives the photon flux $\bar{f}_P$

$$\bar{f}_P = \bar{C}_P/A_{eff}(\theta,\phi), \qquad (17)$$

for the energy range 50–300 keV, where $A_{eff}(\theta,\phi)$ is the effective area of the detector for the direction $\theta,\phi$ of the bursts. Clearly the flux (versus count rate) threshold for a burst in a given direction is also given by a similar expression $\bar{f}_{lim} = \bar{C}_{lim}/A_{eff}(\theta,\phi)$ (see Caditz 1995). Since in our method we directly take into account the truncation $\bar{C}_P \leq \bar{C}_{lim}$ we can deal with the physically more meaningful fluxes which will be subject to the same kind of truncation $\bar{f}_P \geq \bar{f}_{lim}$.

Figure 1 shows this bivariate distribution, which differs from the $\bar{C}_P$–$\bar{C}_{lim}$ distribution by having the points moved diagonally parallel to the truncation line by a factor of $A_{eff}^{-1}$ on both axes. The use of fluxes rather than counts naturally increases the range and dispersion of the variables, but ensures that we are dealing with the more physical quantities.

First we test the data for a correlation between $\bar{f}_P$ and $\bar{f}_{lim}$. We expect that the threshold flux (determined from the background flux before a burst) will have no relation to the peak flux of a subsequent burst. Indeed our analysis using the method described in § 2.3 shows no correlation, with the test results shown as the first entry in Table 1. This method (see Efron & Petrosian 1992 for complete details) evaluates a scalar test statistic $t_w$, the magnitude of which is directly related to the probability $P(t_w) = \mathrm{erfc}(t_w/\sqrt{2})$ that the observed data was drawn from an uncorrelated population. If $|t_w| < 1.645$, then the probability that the data was drawn from an uncorrelated population exceeds 10%, and one can say that the data show no evidence for correlation at the 90% confidence level. The value of $t_w$ we obtain for the 3B sample is 0.78, which corresponds to $P(0.78) = 43.5\%$, confirming our expectation of no correlation.

Next we need to determine the correlation between flux and duration from the trivariate distribution of $\bar{f}_P$, $\bar{f}_{lim}$, and $T_{90}$. A two-dimensional representation of this distribution is presented in Figure 3, where the distribution of bursts projected into the $T_{90}$–$\bar{f}_P$ plane is shown with different plot symbols represent the different values of $\bar{f}_{lim}$. As described in §3.1, there is a bias against the detection of short duration ($T_{90} < \Delta t$) bursts. This bias is evident from the paucity of points in the upper left portion of the figure. Bursts which should have occupied this portion of the figure have moved down to the lower left portion, and bursts which should have occupied the lower left portion have moved below the threshold rate and are not observed. This bias therefore tends to introduce an artificial positive correlation between $T_{90}$ and $\bar{f}_P$.

Determination of the correlation between $T_{90}$ and $\bar{f}_P$ for the whole sample of bursts is complicated because of the variability of $\bar{f}_{lim}$ and the three dimensional nature of the data. However, if we limit ourselves to the subset of bursts with values of $\bar{f}_P$ larger than the largest value of $\bar{f}_{lim}$ (bursts above the solid line in Fig. 3) the problem is reduced to two dimensions. Following the procedure described above, we find that there is indeed a significant correlation between $T_{90}$ and $\bar{f}_{lim}$ (second entry in Table 1).

However, we are interested in the correlation between the duration and the true peak flux $f_P$. For this we use equation (16) to obtain our estimate of this true $f_P$:

$$f_P = \bar{f}_P \left( \frac{T_{90} + \Delta t}{T_{90}} \right). \tag{18}$$

Note that this correction can be applied to the $f_{P,1024}$ but not to the $f_{P,64}$ because we have no measurements at trigger times $\Delta t < 64$ ms. We must now determine the correlation between $T_{90}$ and $f_P$ and their distributions from the data set consisting of $f_P$, $\bar{f}_{lim}$, and $T_{90}$. Figure 4 shows a two-dimensional representation of this data. The solid line, which is the horizontal line in Figure 3 transformed according to equation (18), shows that the short duration bursts with low values of $f_P$ are truncated. Using the methods described in connection with Figure 1 we can test for the independence of $f_P$ and $T_{90}$ using this truncated data. Again limiting our analysis to the two-dimensional case and to the sample of 318 bursts above the truncation line we find only marginally significant evidence for dependence (a small positive correlation indicated the third entry in Table 1). We see no sign of the anticorrelation expected for cosmological time dilation. This aspect is more completely described in another publication (Lee & Petrosian 1996a).

In this paper we are concerned with the determination of bias-free univariate distributions of the true peak fluxes and durations. It is desirable to use the maximum possible number of data points so that consideration of the three-dimensional nature of the distribution becomes necessary.

The three-dimensional data can be converted to the generic two-dimensional case by the transformation of the variable $T_{90}$ to a new variable, which we call $f_{lim}(T_{90})$, given by an expression identical to equation (18):

$$f_{lim}(T_{90}) = \bar{f}_{lim} \left( \frac{T_{90} + \Delta t}{T_{90}} \right). \tag{19}$$

Note that $f_{lim}(T_{90})$ should be considered a function of $T_{90}$ and not a true threshold. The instrumental threshold $\bar{f}_{lim}$ is a true threshold and is not expected to be correlated with anything at all. The limiting flux $f_{lim,i}(T_{90,i})$ for burst $i$ is the minimum value its true peak flux $f_{P,i}$ must have had so that it would have triggered under the criterion $\bar{f}_{P,i} > \bar{f}_{lim,i}$. With this transformation we have now a bivariate distribution of $f_P$ and $f_{lim}(T_{90})$, with the generic truncation $f_P > f_{lim}(T_{90})$. Figure 5 shows this distribution. Just as in the transformation from $(\bar{C}_P, \bar{C}_{lim})$ to $(\bar{f}_P, \bar{f}_{lim})$, the transformation from $(\bar{f}_P, \bar{f}_{lim})$ to $(f_P, f_{lim}(T_{90}))$ amounts to sliding the data points parallel to the truncation line, this time by an amount equal to $1 + \Delta t/T_{90}$. Clearly, long duration bursts ($T_{90} > \Delta t$) are not affected by this transformation, but for short duration bursts $f_P$ and $f_{lim}(T_{90})$ increase, which increases the range of the variables. Even though $\bar{f}_P$ and $\bar{f}_{lim}$ are independent, $f_P$ and $f_{lim}(T_{90})$ may be correlated. Such a correlation would reflect any correlation that may exist between $f_P$ and $T_{90}$. As shown by the fourth entry in Table 1, we find a statistically insignificant anticorrelation between $f_P$ and $f_{lim}(T_{90})$ consistent with the small correlation found above between $f_P$ and $T_{90}$ using the limited sample. This result will be used in the determination of the univariate distribution of the true peak flux $f_P$.

Before moving on, we note that a similar transformation may be performed on the data to aid determination of the distribution of $T_{90}$. Instead of transforming the duration, we transform the peak flux via the inverse of the transformation given by equation (19),

$$T_{lim}(f_P) = \Delta t(f_P/\bar{f}_{lim} - 1)^{-1}, \tag{20}$$

so that the problem is again reduced to the generic case of a bivariate distribution with $T_{90} > T_{lim}(f_P)$. Again the limiting duration $T_{lim,i}(f_{P,i})$ of burst $i$ is the minimum value its duration $T_{90,i}$ must have had so that it would have been triggered, i.e. it would have $\bar{f}_{P,i} > \bar{f}_{lim,i}$. The $T_{90}$–$T_{lim}$ distribution is shown in Figure 6. The result of the independence test (fifth entry in Table 1) shows an anticorrelation between $T_{90}$ and $T_{lim}(f_P)$, which in its sense is consistent with the above results. However, the strength of the correlation is slightly stronger than what was found in the previous tests. The difference in strength may be due to a weak but complicated correlation between $f_P$ and $T_{90}$, so that the correlation between these quantities is affected by the transformations given by equations (19) and (20). In particular, the range of $T_{lim}(f_P)$ for bursts with $T_{90} > \Delta t$ is very sensitive to the functional form chosen in equation (16) and utilized in equations (19) and (20). For example, if we had chosen the form appropriate for the square pulse (see Fig. 2), all the points in Figure 6 with $T_{lim}(f_P) > \Delta t$ would collapse on the line $T_{lim}(f_P) = \Delta t$, while the points in Figure 5 would be essentially unaffected.

We therefore conclude that there seems to be a marginally significant correlation between $f_P$ and $T_{90}$ which most likely should not greatly affect the determination of the univariate distributions.

### 3.3. Distribution of Peak Fluxes

From the various bivariate distributions described above we can now find univariate distributions. We first discuss the distribution of peak fluxes. We present not only the cumulative distribution customarily used in the analysis of GRB data, but also a useful form of the differential distribution and the variation of logarithmic slope defined in equation (4). As is the practice in the analysis of radio source counts, we do not present the differential distribution $n(f_P)$ defined by equation (1); instead, we present $f_P^{5/2} n(f_P)$, which shows the deviation from HISE, in which the expected distribution is proportional to $f_P^{-5/2}$. Figure 7 shows the results. In each case the solid lines show the distributions of the peak fluxes $f_P$ without consideration of the truncations. The dotted histograms show when the effects of the duration bias and variable threshold are included via application of the method associated with Figure 5.

Note that, as expected, the truncations tend to discriminate against the detection of weaker (lower peak flux) bursts, so that when properly accounted for the numbers of weak bursts is increased and the slope becomes larger (steeper $N(f_P)$ or $n(f_P)$). We have not included the effect of the small correlation which we found in the $f_P$–$f_{lim}(T_{90})$ distribution of Figure 5. Inclusion of this effect adds a qualitatively similar but quantitatively insignificant correction. The steepening of the corrected distributions must be taken into account when comparing the data with models.

### 3.4. Distribution of Durations

The first analysis of the duration distribution of BATSE data was carried out by Kouvelioutou et al. (1993), who plotted distributions for all bursts with known values of $T_{90}$ and $T_{50}$ in the 1B catalog. These distributions showed the possible existence of two populations of bursts: short duration and long duration, with the division at $T \approx 2$ s. This analysis was done without regard for the various data truncations, nor were the effects of possible correlations between peak count rate $C_P$ and duration acknowledged or considered. These truncations and correlations affect the distribution of durations as they do the distribution of fluxes. In paper PII we showed that when the first two effects are properly taken into account the number of bursts with short durations is enhanced considerably, making the distinction between the two populations even more striking. We have repeated this analysis for the 3B catalog, now also including the effects of the correlation between flux and duration. Figure 8 shows our results. It turns out that the new bursts in the 3B-1B catalog, when limited to bursts with $\bar{C}_P > \bar{C}_{lim}$, do not show as strong evidence for two populations as the 1B, so that in the total 3B catalog this distinction is somewhat diminished and is barely visible in the raw data, shown as the solid histogram.

However, when we include the effect of the duration bias, we obtain the dotted histogram which clearly shows a significantly larger population below $T_{90} < 2$ s. Recall that in the analysis of this data we found a stronger correlation between $T_{lim}(f_P)$ and $T_{90}$ than we would have expected considering the weak correlation between their counterparts $f_P \propto (1 + \Delta t/T_{90})$ and $f_{lim}(T_{90}) \propto (1 + \Delta t/T_{90})$. We suspect this difference is due to the form of these transformations, where small fluctuations in the fluxes can produce large fluctuations in the durations when the latter are greater than $\Delta t = 1.024$ s. Nevertheless, since the $T_{90}$–$T_{lim}(f_P)$ distribution shown in Figure 6 shows a significant correlation, the simple application of the methods of §3.3 will not be correct. We must find a set of variables which are uncorrelated. Unfortunately this step cannot be carried out non-parametrically and we must assume a correlation form (i.e. a parametric transformation of the variables). A simple description of this process can be found in Lee et al. (1993 and 1995). Since we are interested in the distribution of $T_{90}$ we transform the variable $T_{lim}$ to a new one $T'_{lim} = T_{lim} T_{90}^{\alpha_T}$ and determine the correlation test result $t_w$ as a function of $\alpha_T$. We take the value of $\alpha_T = 0.18^{+0.12}_{-0.11}$ for which $t_w = 0 \pm 1.645$ as our best estimate for the degree of correlation between $T_{90}$ and $T_{lim}$. Using $\alpha_T = 0.18$, we transform into the $T_{90}$–$T'_{lim}$ plane. Since the variables are now uncorrelated we apply the method of §3.3 to obtain the distribution shown by the dashed histogram of Figure 8, which shows an even stronger peak in the distribution at short durations.

As mentioned above, because of the uncertainty in the transformation this last result is subject to error. To determine the degree of uncertainty we have used two other transformations, one appropriate for a square pulse (Eq. [13]) and one for a spiky pulse $f(t) = f_P(1 - |t/T|^n)$ (note that the case $n = 1$ applies for a triangular pulse). We find that for the square pulse, $t_w = -3.36$, while for the spiky pulses $t_w = -1.22, -1.58, -1.77$ for $n = 1, 2, 3$. The calculated correlations as measured by $\alpha_T$ vary by up to a factor of two from the value calculated above which was used to produce the figure, leading to distributions that are uncertain by this amount at short durations. Although the magnitude of the correction is fairly uncertain, application of the correlation correction can only increase the number of short duration bursts relative to the uncorrected case.

## 4. Discussion and Summary

None of our conclusions change significantly if we use $T_{50}$ rather than $T_{90}$ as our estimate of the duration, or if we use one of the other time scales (64 ms or 256 ms) for $f_P$ (note that for $\Delta t = 64$ ms we cannot test the validity of Eq. [16] because there exists no time scale below 64 ms). We have chosen to present the 1024 ms results because the trigger sensitivity goes as $\sqrt{\Delta t}$, resulting in a larger sample. In this study we have used $T_{90}$ rather than $T_{50}$ so that our results may be directly compared with previous studies (e.g. Kouveliotou et al. 1993, Norris et al. 1994), which typically use $T_{90}$ or some variation of it. In addition, as a third measure of duration we take advantage of the fact that the energy fluences $\mathcal{F}_E$ are tabulated in the BATSE catalog. We define the effective duration

$$T_{eff} = \frac{\mathcal{F}_E}{f_P \langle h\nu \rangle}, \tag{21}$$

where $\langle h\nu \rangle$ is the average energy per photon, which for each burst can be deduced from the published hardness ratios with an assumption about the form of the spectra (we assumed that they could be described as power laws). Qualitatively, the distributions derived using $T_{eff}$ or $T_{50}$ are very similar to those derived using $T_{90}$. One of the few notable differences is that the univariate $T_{eff}$ and $T_{50}$ distributions corresponding to Figure 8 extend over slightly less dynamic range, as expected due to the definitions of these durations. In addition, the correlations are somewhat weaker when using these measures of duration instead of $T_{90}$. The implications of these small differences are discussed more fully in another work (Lee & Petrosian 1996a).

Our results show the presence of a significantly larger population of short duration bursts than observed by BATSE. The question then arises if there exist two distinct populations of GRBs. As further evidence for this we present Figure 9 in which we plot the individual cumulative peak flux distributions for 8 different duration groups, each containing 65 sources. There does not appear to be any systematic variation in slope for the top six duration bins (confirming the results of § 3.2), but there seems to be a significant increase in the slope of these histograms for the last two bins where $T_{90} < 4$ s. In particular, the value of the slope of the last bin ($T_{90} < 1$ s), $-1.3$, is not significantly different from the value of $-1.5$ expected from HISE. One possible explanation for this difference would be that if there is indeed a real flattening of the distribution at low peak fluxes, the duration bias (which increases with decreasing duration) reduces the observed fluxes enough to truncate those points that would have shown the flattening. However, for the top seven duration bins combined, the slope of the distribution of $f_{P,1024}$ with the data truncated as $f_{P,1024} > 1$ ph cm$^{-2}$ s$^{-1}$ (approximately the range covered by the eighth duration bin) is $-0.99$.

The same difference in the last two bins is also present in the distributions of $\bar{f}_P$ and $\bar{C}_P$. However, as mentioned in § 1, for short duration GRBs (those with $T \leq \Delta t$), the triggering of BATSE and other instruments is based on the photon fluence $\mathcal{F}$, not the peak photon count rate. This fluence must exceed the threshold $\mathcal{F}_{lim} = \bar{f}_{lim} \Delta t$. Therefore, for these bursts the only bias introduced is due to the variability of $\mathcal{F}_{lim}$ so that we can obtain a reliable $\log N$-$\log \mathcal{F}$ histogram, which of course would be identical in shape to the $\log N$-$\log \bar{f}_P$ histogram since $\bar{f}_P = \mathcal{F}/\Delta t$. The intriguing result presented above is that for such bursts where the complication due to light curves plays no role the analysis does not show a very significant deviation from HISE.

This result could lead us to believe that perhaps the peak flux $f_P$ or photon count rate $C_P$ is not a good measure of the basic energy release (and therefore distance) of GRBs. Indeed, the wide dispersion in duration and the multiplicity and complexity of light curve shapes make it unlikely that the peak of the largest spike could have much physical significance. Other characteristics such as the fluence of individual spikes, the total fluence, the duration of individual spikes, or the number and separation of the spikes may be more robust characteristics. If fluence turns out to be such a characteristic, then our result about the distribution of fluence of short duration bursts indicates that there exists the distinct possibility that short duration bursts are weak and nearby (distance less than several hundred parsecs) and therefore very nearly isotropic. It would be helpful if similar robust results can be obtained for longer duration bursts. We defer further discussion of these and other issues concerning the fluence to another publication (Lee & Petrosian 1996b).

To summarize, the results of this paper are:

1. We have emphasized that the data describing GRBs forms a multivariate distribution. Typically, it is desirable to extract univariate distributions from the multivariate distribution in order to compare the observations with theory. The multivariate nature of the distribution is important because the highly variable nature of GRBs and the procedures followed for their detection by instruments such as BATSE lead to complex data truncations in multivariable space. Hence, the distribution of each variable depends on the other variables, so that no single variable can be considered by itself. If the truncation effects are not properly accounted for, biased and incorrect univariate distributions can result.

2. Biases in the distributions can be a result of not only data truncation, but also correlations between the variables in the multivariate distribution. We have presented methods to extract univariate distributions in the presence of both truncation and correlations.

3. We have applied these methods to the BATSE 3B data set to extract the distribution of peak flux (the so-called $\log N$-$\log S$ relation) and duration of the bursts from the multivariate distribution. We find that there is little evidence for any correlation between the flux and duration. If anything, the correlation is in the opposite sense of that required by cosmological time dilation and does not support the conclusions of Norris et al. (1994 & 1995). When corrected for the various truncations and correlations, the duration distribution shows many more short duration events than shown in the distribution published by Kouveliolutou et al. (1993).

4. We show that the $\log N$-$\log S$ relation gets increasingly steeper as the corrections for the various biases are applied. We find that for short duration bursts, the slope of the $\log N$-$\log S$ relation does not differ significantly from that predicted by HISE. We offer a possible explanation that the short duration bursts form a separate population of local isotropically distributed sources.

This work has benefited from discussions with B. Efron and C. Meegan. We also acknowledge support from NASA grants NAGW 2290 and NAG-5 2733.

## A. Correction for the Peak Flux Bias

The use of $\bar{C}_{P,64}$ and $\bar{C}_{P,1024}$ as estimates of the peak count rate in the intervals $\Delta t = 64$ ms and 1024 ms respectively introduces the peak flux bias: $\bar{C}_{P,\Delta t}$ is an upwardly biased estimator of the true peak count rate in the interval $\Delta t$. This bias can be understood by considering the limiting case of a burst with a flat time profile of count rate $C$ and duration $T$. The measured $\bar{C}_{P,\Delta t}$ will differ from $\Delta t C$ because of Poisson fluctuations inherent in the measurements. Since $\bar{C}_{P,\Delta t}$ is defined to be the *maximum* number of counts in the interval $\Delta t$, for $T \gg \Delta t$ it is likely that these rates will exceed $C\Delta t$ because the distribution of $\bar{C}_{P,\Delta t}$ has a finite spread about $C\Delta t$. Note that $\bar{C}_{P,64}$ need not occur within the interval during which $\bar{C}_{P,1024}$ occurs.

### A.1. Estimates of the Bias

We may estimate the effect of the peak flux bias on the ratio $v = \bar{C}_{P,64}/\bar{C}_{P,1024}$ as follows. For $T < 1024$ ms, the ratio of the number of samples at 64 ms to the number of samples at 1024 ms (1 sample) increases with increasing $T$, so that the bias should be negligible at $T < 64$ ms and should increase rapidly with $T$. For $T \gg 1024$ ms, the bias should increase very slowly with $T$ because the ratio of the number of samples at 64 ms to the number of samples at 1024 ms is approximately constant ($\approx 16$). The bias should depend not only on $T$ but also on $C$, with weak bursts being more affected by the bias because Poisson fluctuations in the background are relatively more important for these bursts. If we consider a constant count rate burst with count rate $C$, background count rate $B$, and duration $T$, sampled at intervals $\Delta t$, the number of samples is given by $N = T/\Delta t$ and the true maximum total counts is $\mu = (C+B)\Delta t$. Because the actual measured counts obey Poisson statistics, there will be some spread in the distribution of the $N$ measured counts so that the maximum counts likely will exceed $\mu$. The distribution of measured counts $c$ is given by $P(c,\mu)T/\Delta t$, where $P$ is the Poisson distribution. The maximum expected measured counts $c_{max}$ is given by

$$1 = \int_{c_{max}}^{\infty} P(c,\mu)dc. \tag{A1}$$

Because in general $\mu \gg 1$ (backgrounds alone are on the order of $10^2$ counts in 64 ms), to a good approximation a Gaussian distribution may be substituted for the Poisson distribution, so that

$$\frac{\Delta t}{T} = \frac{1}{\sqrt{\pi}} \int_{(c_{max}-\mu)/\sigma}^{\infty} \exp(-u^2)du, \tag{A2}$$

where $\sigma = \sqrt{\mu}$, or

$$\frac{2\Delta t}{T} = \mathrm{erfc}\left(\frac{c_{max} - \mu}{\sigma}\right), \tag{A3}$$

so that

$$c_{max} = \underbrace{(C+B)\Delta t}_{\mu} + \underbrace{\sqrt{(C+B)\Delta t}\,\mathrm{erfc}^{-1}\left(\frac{2\Delta t}{T}\right)}_{\text{excess due to bias}}. \tag{A4}$$

The ratio $v = \bar{C}_{P,64}/\bar{C}_{P,1024}$ is thus given by

$$\begin{aligned} v &= \frac{1.024}{0.064} \cdot \frac{c_{max}(\Delta t = 0.064) - 0.064 B}{c_{max}(\Delta t = 1.024) - 1.024 B} \\ &= 16 \frac{0.064 C + \sqrt{0.064(C+B)}\,\mathrm{erfc}^{-1}\left(\frac{0.128}{T}\right)}{1.024 C + \sqrt{1.024(C+B)}\,\mathrm{erfc}^{-1}\left(\frac{2.048}{T}\right)}. \end{aligned} \tag{A5}$$

It can be seen that $v$ exceeds $v_{true} = 1$ because of the excess in both the numerator and the denominator due to the bias. The ratio of $v$ to $v_{true}$ depends on $T$, $C$, and $B$. For a given background rate and duration, it is evident that the larger bursts will be less affected by the excess due to the bias, because for $C \to \infty$, $v \to v_{true}$. Large backgrounds tend to increase the bias, as do long durations. Table 2 lists some values of $v/v_{true}$ for several representative values of $C$, $T$, and $B$. It can be seen that the corrections due to the bias range from a few percent to up to a factor of two, depending on the parameters.

## A.2. Bias Simulation

To investigate the peak flux bias more thoroughly, we performed a simulation to find the ratio $v = \bar{C}_{P,64}/\bar{C}_{P,1024}$ for constant count rate bursts of varying $C$ and $T$. The bias is maximized for bursts of constant count rate and high background, so we chose an expected background count rate of $B = 4000$ counts/sec, which is close to the maximum background rate listed in the BATSE catalog. For a given $C$ and $T > \Delta t$, we Poisson deviate the expected counts $\Delta t(C + B)$ for each of $n$ intervals of time $\Delta t$ until $(n+1)\Delta t > T$. For instances where 64 ms $< T <$ 1024 ms (or when $\Delta t$ does not divide evenly into $T$), there will be $m$ 64 ms intervals with zero expected excess counts. For these $m$ intervals, we Poisson deviate the background counts. From the $n + m$ intervals, the one with the largest number of counts is defined to be $(\bar{C}_{P,\Delta t} + B)\Delta t$ and $\bar{C}_{P,\Delta t}$ is readily found and the ratio $v$ is found as a function of $\bar{C}_{P,1024}$ and $T$.

To approximately correct the ratio $v$ derived from the BATSE data for the peak flux bias, we use the results of our simulation. For each burst, we use $\bar{C}_{P,1024}$ and $T_{90}$ to find the associated $v$ in our simulation. Since we know what $v_{true}$ is for a given $T$, we use the results of the simulation to map the observed $v$ to $v_{true}$. We note that strictly it is only possible to map forward from $v_{true}$ to $v$ due to the variance caused by the Poisson fluctuations; thus a backward mapping from $v$ to $v_{true}$ will only be approximately correct. Since we are interested not in accurately deriving $v_{true}$ for any individual burst but rather in determining gross changes in $v_{true}$ with $T_{90}$, this approximation is reasonable for our purposes. We also note that we are assuming the maximal bias by using constant count rate bursts and a large background in our simulation.

The results of this correction are shown in Figure 2. It can be seen that:

1. As expected, the bias increases rapidly with $T_{90}$ until $T_{90} \gtrsim 1$s.

2. The maximum effect of the bias on $v$ is approximately a factor of two as predicted in §A.1.

3. The correction for the bias reduces the average value of $v$ to a value slightly less than 1 at large $T_{90}$, indicating that the effect of the peak flux bias has been overestimated as predicted. We note that the corrected $v$ versus $T_{90}$ curve fits the curve given by equation (16) remarkably well.

| Test | Burst Sample | $t_w$ | $P(t_w)$ |
|---|---|---|---|
| $\bar{f}_P$ vs. $\bar{f}_{lim}$ | all available (518 bursts) | 0.78 | 0.44 |
| $\bar{f}_P$ vs. $T_{90}$ | max sample with constant $\bar{f}_{lim}$ (385 bursts) | 4.17 | $3.0 \times 10^{-5}$ |
| $f_P$ vs. $T_{90}$ | max sample with constant $\bar{f}_{lim}$ (385 bursts) | 2.03 | $4.2 \times 10^{-2}$ |
| $f_P$ vs. $f_{lim}(T_{90})$ | all available (518 bursts) | -1.49 | 0.14 |
| $T_{90}$ vs. $T_{lim}(f_P)$ | all available (518 bursts) | -2.76 | $5.8 \times 10^{-3}$ |

Table 1: Correlation test results for various pairs of variables. $P(t_w)$ indicates the probability of obtaining the given value of $t_w$ from the distribution assuming uncorrelated variables. The sign of $t_w$ indicates the sense of the correlation (positive indicates correlation, negative indicates anticorrelation).

| $C$ | $T$ | $B$ | $v/v_{true}$ |
|---|---|---|---|
| (cts/sec) | (sec) | (cts/sec) | |
| 1000 | 100 | 4000 | 1.43 |
| 1000 | 10 | 4000 | 1.40 |
| 1000 | 1000 | 4000 | 1.52 |
| 350 | 100 | 4000 | 1.94 |
| 10000 | 100 | 4000 | 1.08 |
| 1000 | 100 | 2500 | 1.37 |

Table 2: Values of the ratio of observed to true $v = \bar{C}_{P,64}/\bar{C}_{P,1024}$, for various choices of square pulse count rate $C$, duration $T$, and background count rate $B$.

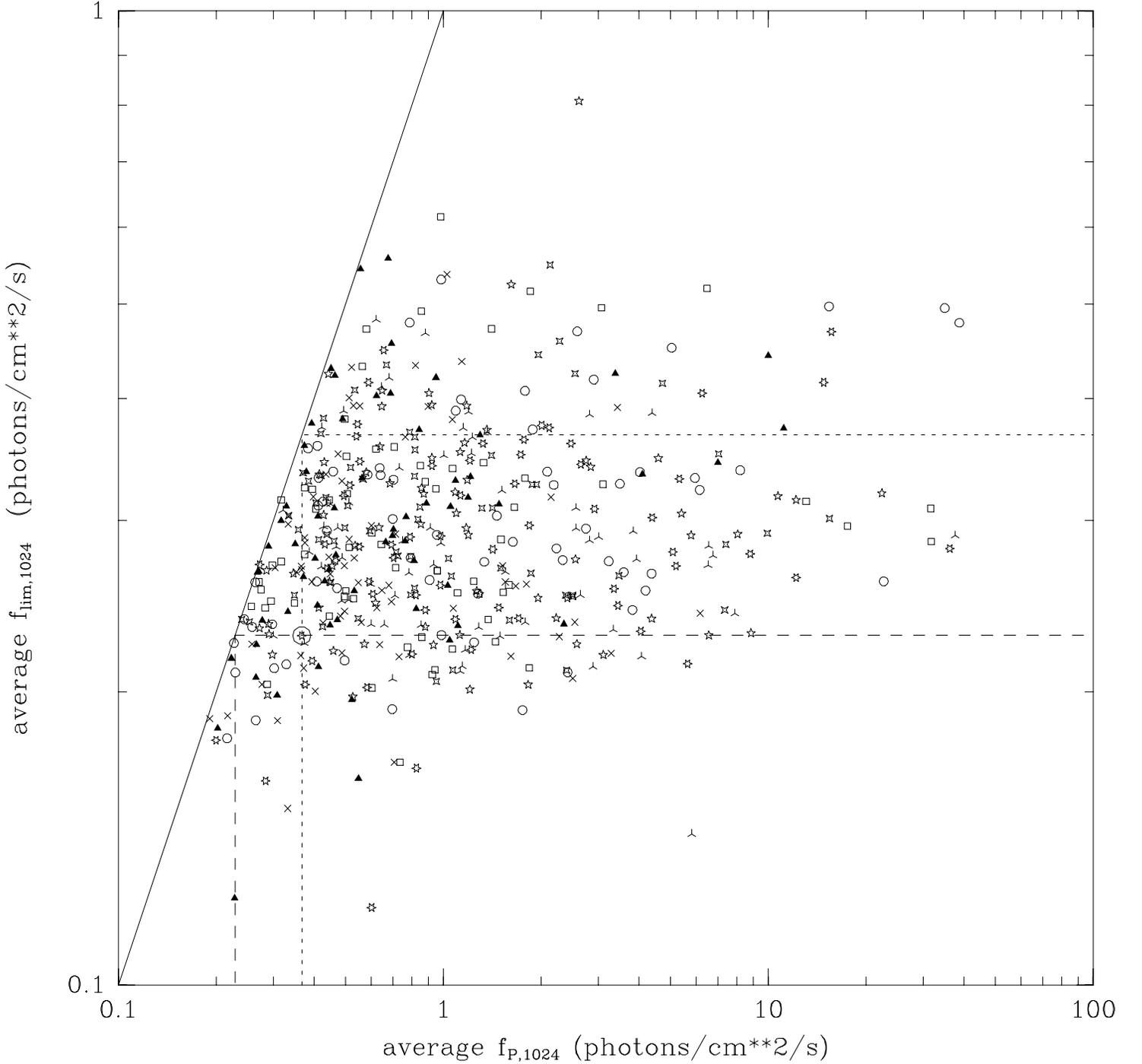

Fig. 1.— The bivariate distribution of $\bar{f}_{P,1024}$ and $\bar{f}_{lim,1024}$. The diagonal line indicates the truncation limit $\bar{f}_{P,1024}/\bar{f}_{lim,1024} > 1$. The boxes enclose the associated sets of a typical data point ($\bar{f}_{P,1024} = 0.367, \bar{f}_{lim,1024} = 0.229$), which is circled in the figure. The short (long) dashed box contains the associated set for the point when considering the distribution along the horizontal (vertical) axis of $\bar{f}_P$ ($\bar{f}_{lim}$). The different plot symbols correspond to bursts of differing $T_{90}$. Each plot symbol represents 12.5% of all bursts. In order of increasing $T_{90}$: crosses, triangles, 4-pointed stars, circles, squares, 5-pointed stars, 3-pointed crosses, 6-pointed stars.

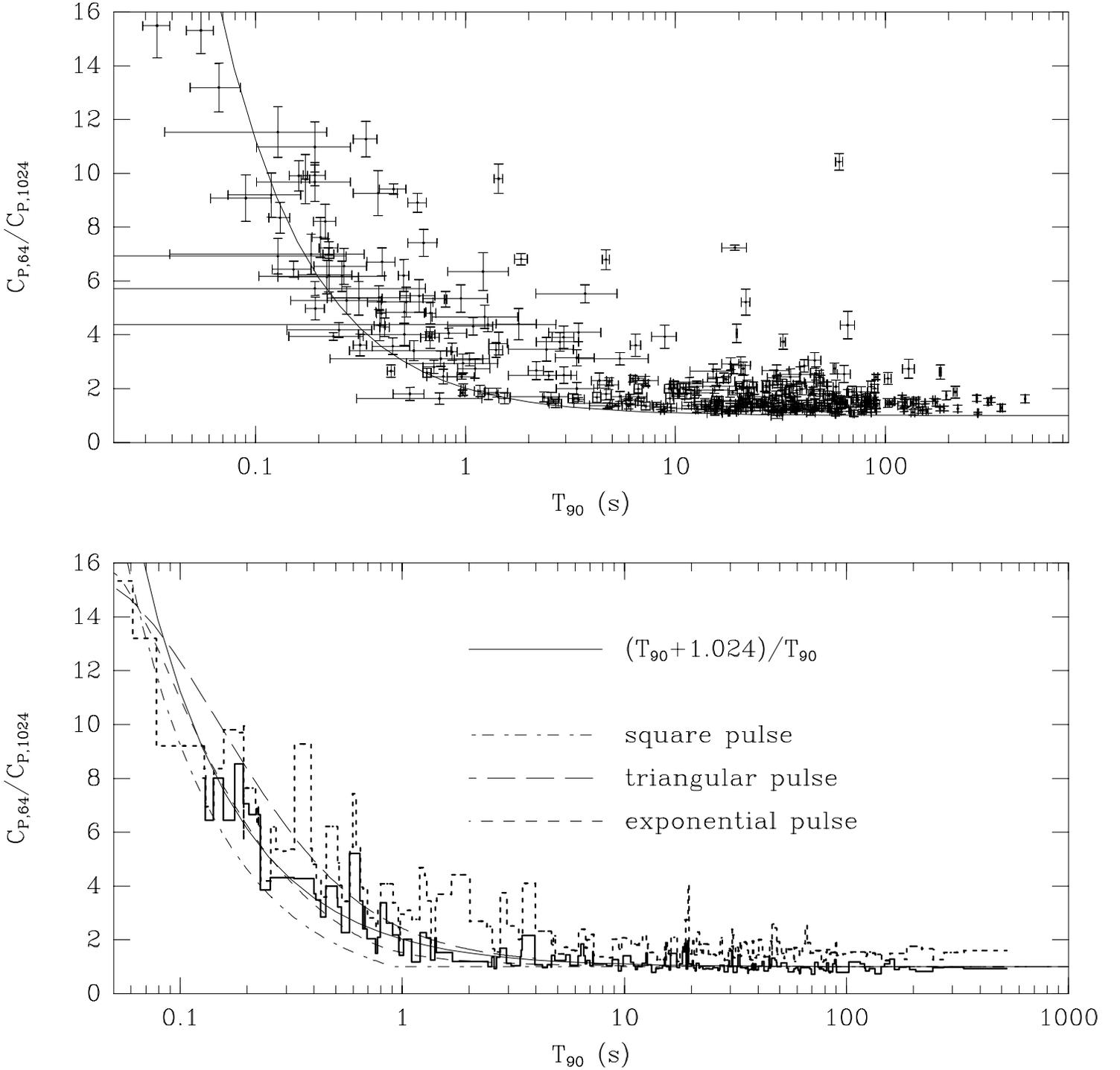

Fig. 2.— *top panel:* The ratio $\bar{C}_{P,64}/\bar{C}_{P,1024}$ for the BATSE 3B data. *bottom panel:* The same ratio, but for clarity the data have been smoothed by a running median filter of 3 points in $T_{90}$. The dotted histogram shows the ratio obtained from the catalog data. The solid histogram shows the ratio corrected for the peak flux bias (see appendix A). The solid histogram does not extend as far as the dotted histogram because we are unable to use our correction procedure for bursts with $T_{90} < 128$ ms. The smooth curves show the theoretical values of this ratio for a variety of pulse shapes, and the approximation of equation (16).

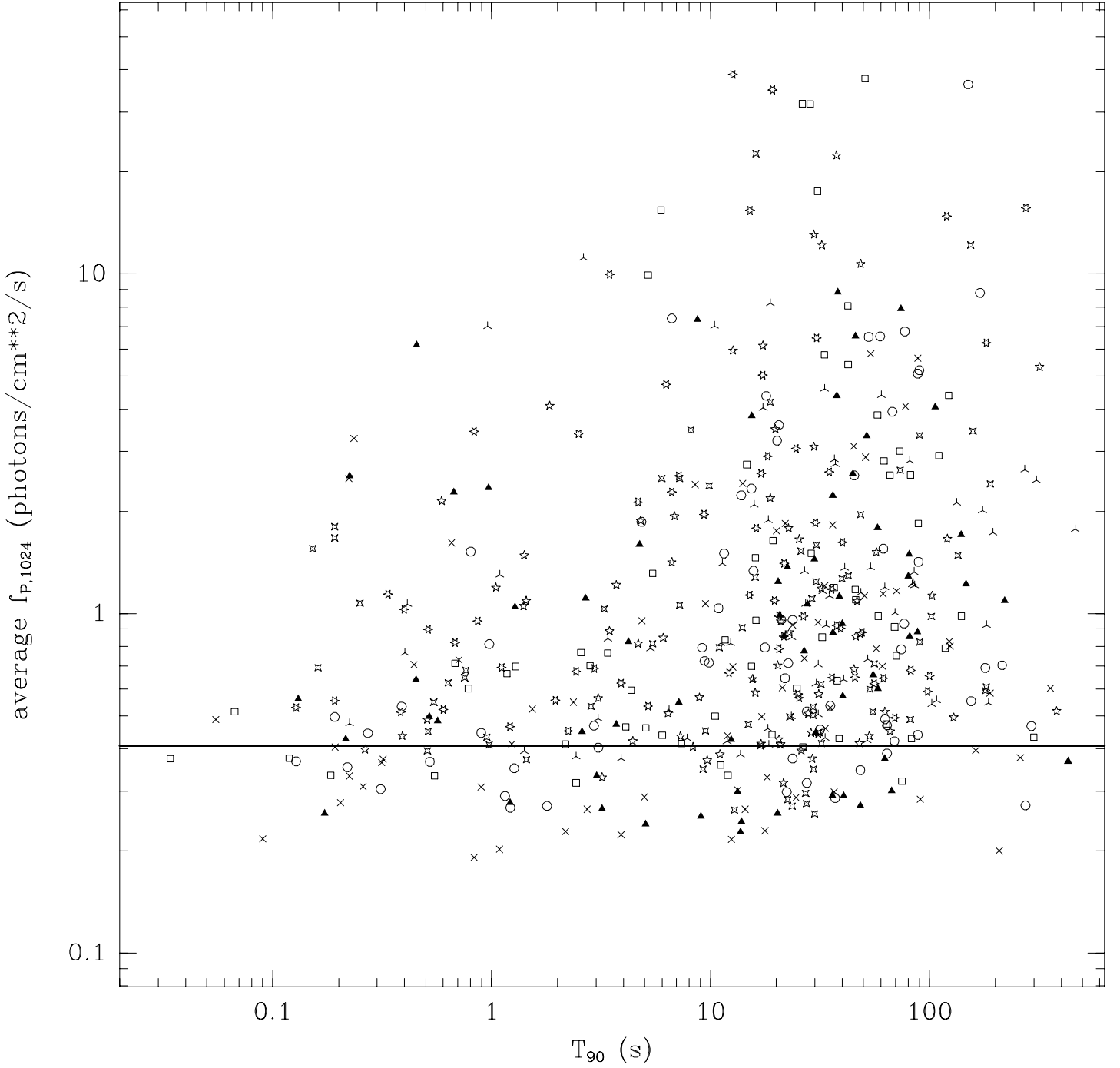

Fig. 3.— The bivariate distribution of average peak flux $\bar{f}_{P,1024}$ and $T_{90}$. The horizontal line indicates the highest level of $\bar{f}_{lim,1024}$, above which the data is complete. The different plot symbols correspond to bursts of differing $\bar{f}_{lim}$. Each plot symbol represents 12.5% of all bursts, with the symbol order the same as in Figure 1.

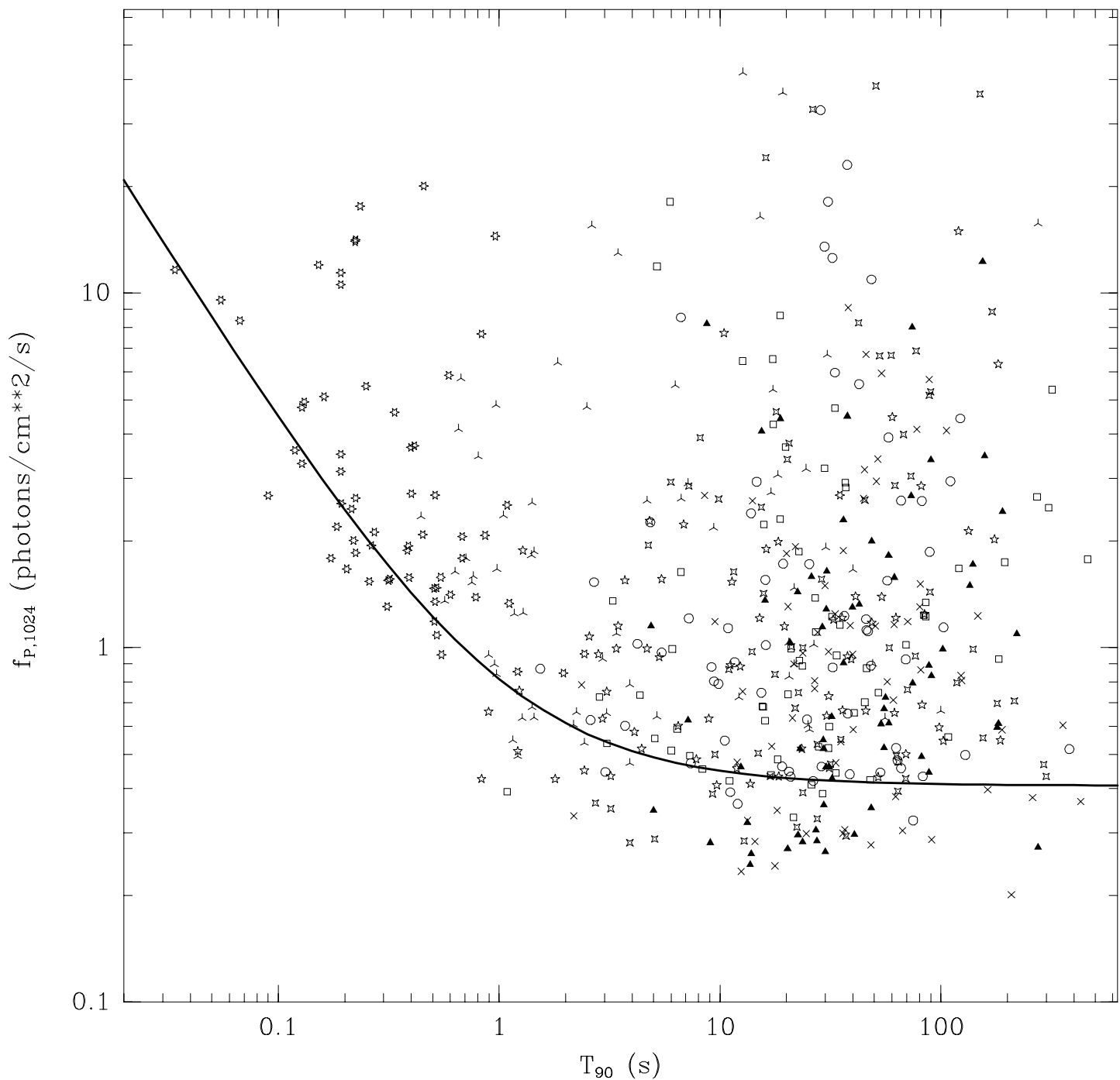

Fig. 4.— The bivariate distribution of true peak flux $f_{P,1024}$ and $T_{90}$. The curve indicates the highest level of $f_{lim,1024}$, above which the data is complete. The different plot symbols correspond to bursts of differing $f_{lim}$. Each plot symbol represents 12.5% of all bursts, with the symbol order the same as in Figure 1.

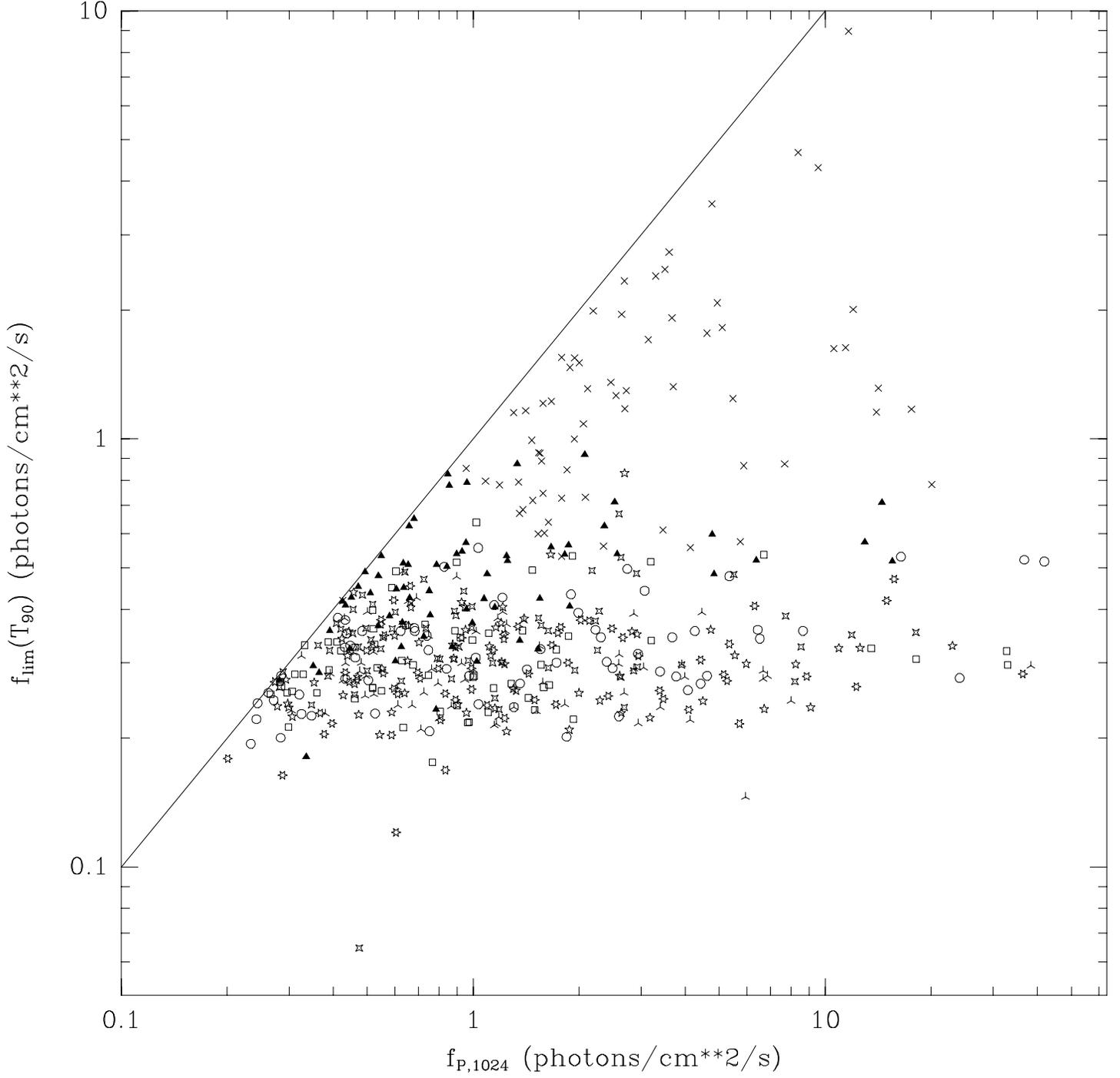

Fig. 5.— The bivariate distribution of $f_P$ and $f_{lim}(T_{90})$. The different plot symbols represent different values of $T_{90}$, with the ordering of the symbols is the same as in Figure 1.

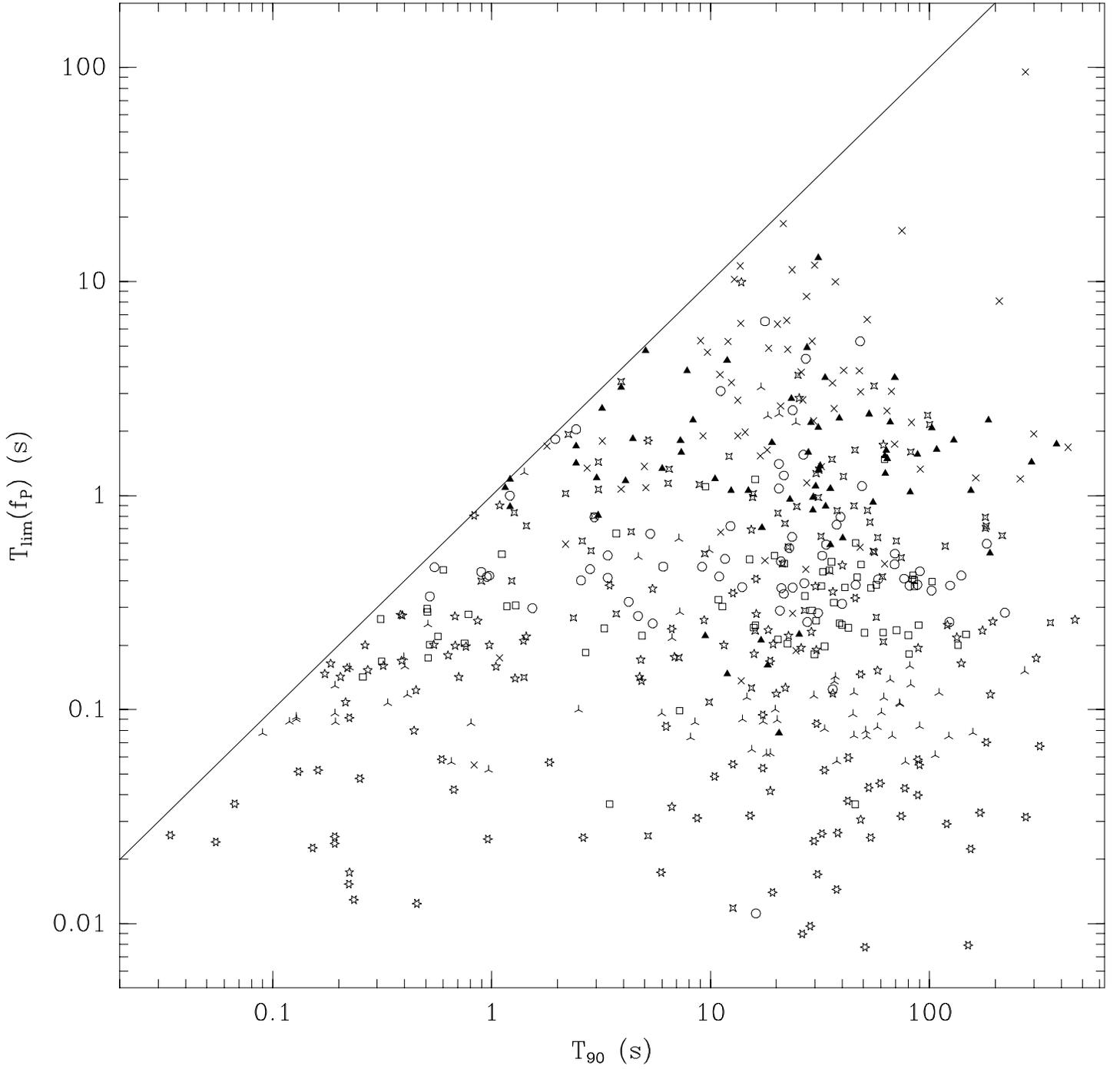

Fig. 6.— The bivariate distribution of $T_{lim}(f_P)$ and $T_{90}$. The different plot symbols stand for different values of $f_P$, with the ordering of the symbols the same as in Figure 1.

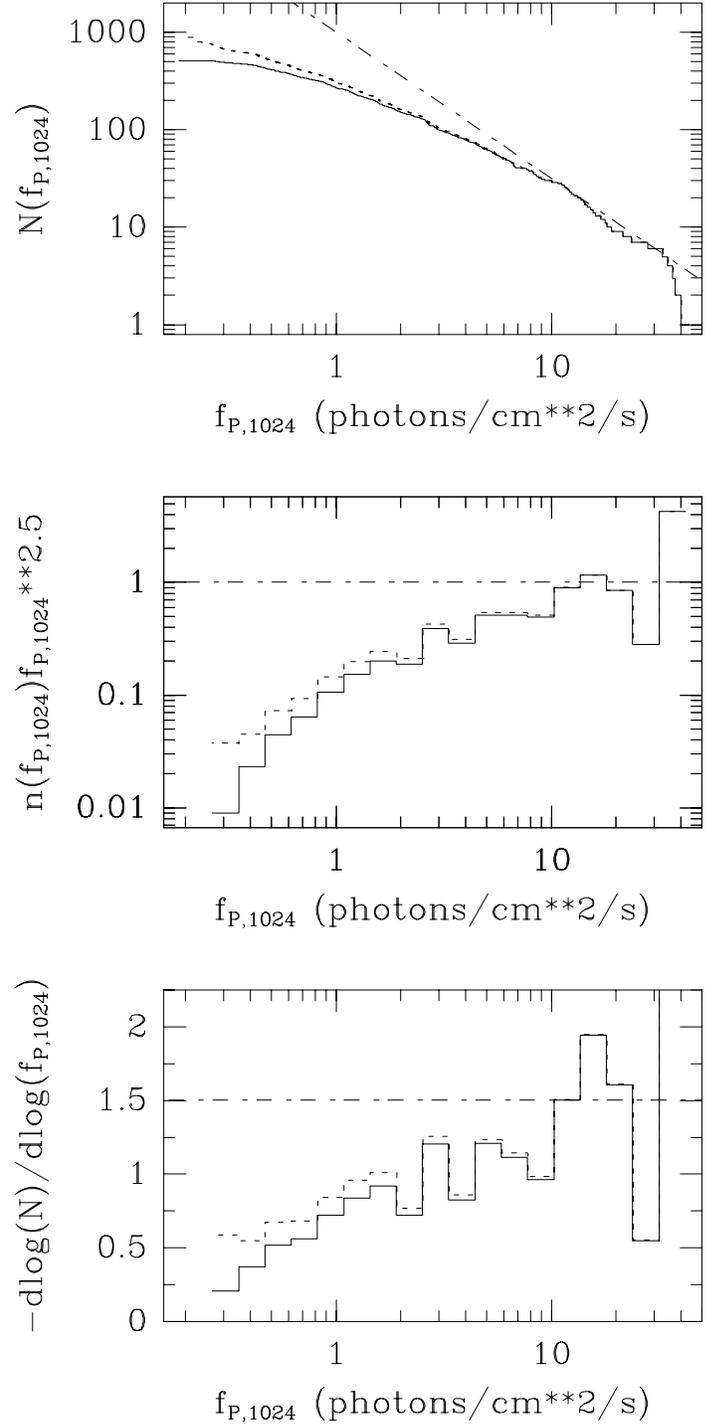

Fig. 7.— Comparison of the cumulative distributions, differential distributions, and logarithmic slopes of the true peak flux (from top to bottom). The solid histograms are what would be found without consideration of the truncation. The dotted histograms are corrected for varying $f_{lim}(T_{90})$. The dot-dashed lines indicate the HISE prediction of logarithmic slopes of -1.5. For the middle graph, the vertical scale is arbitrary.

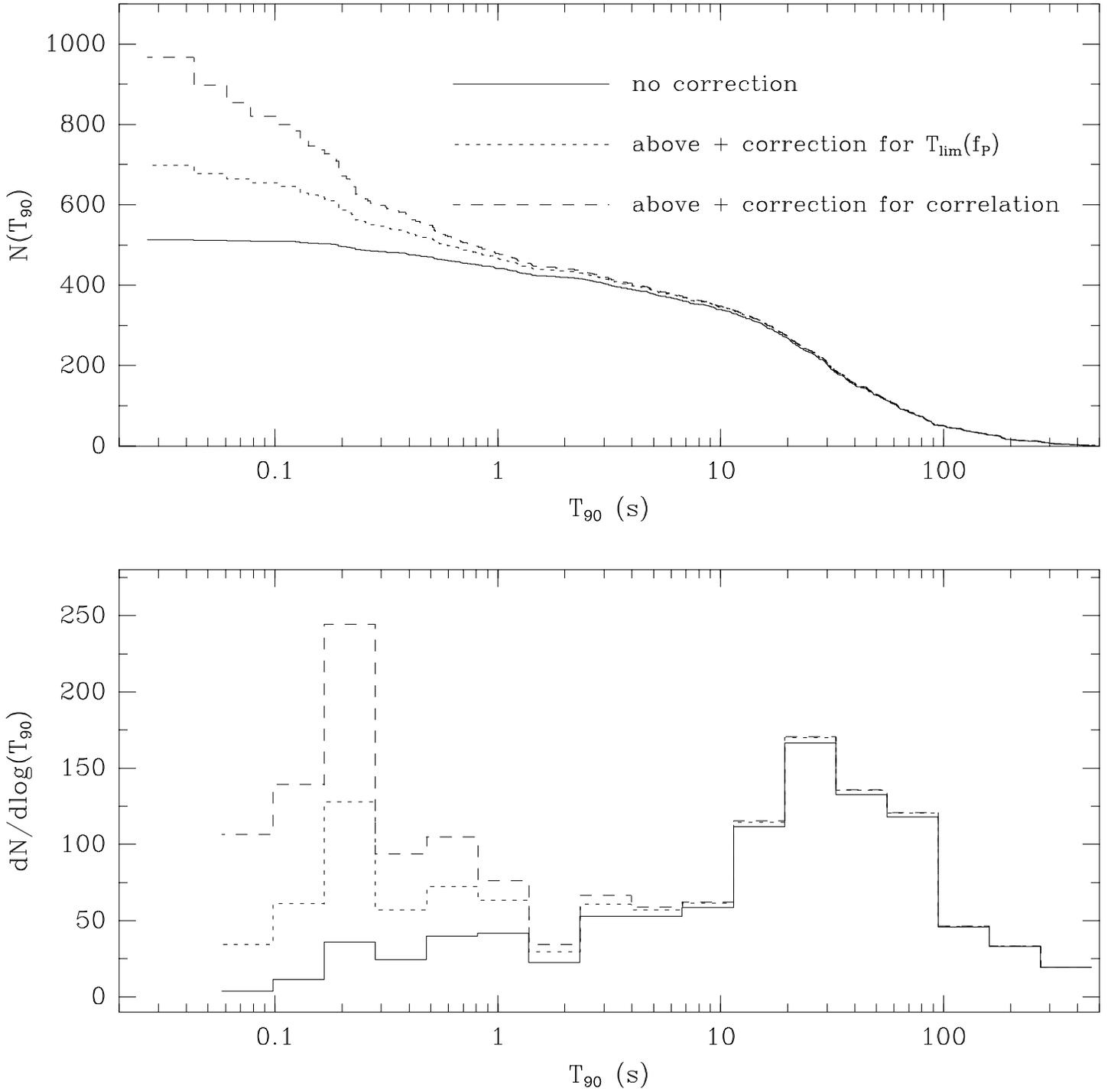

Fig. 8.— Comparison of cumulative and differential distributions of $T_{90}$ (top and bottom panels respectively). The distributions show the effects of the corrections for (from bottom to top) nothing, varying $T_{lim}$, and correlation.

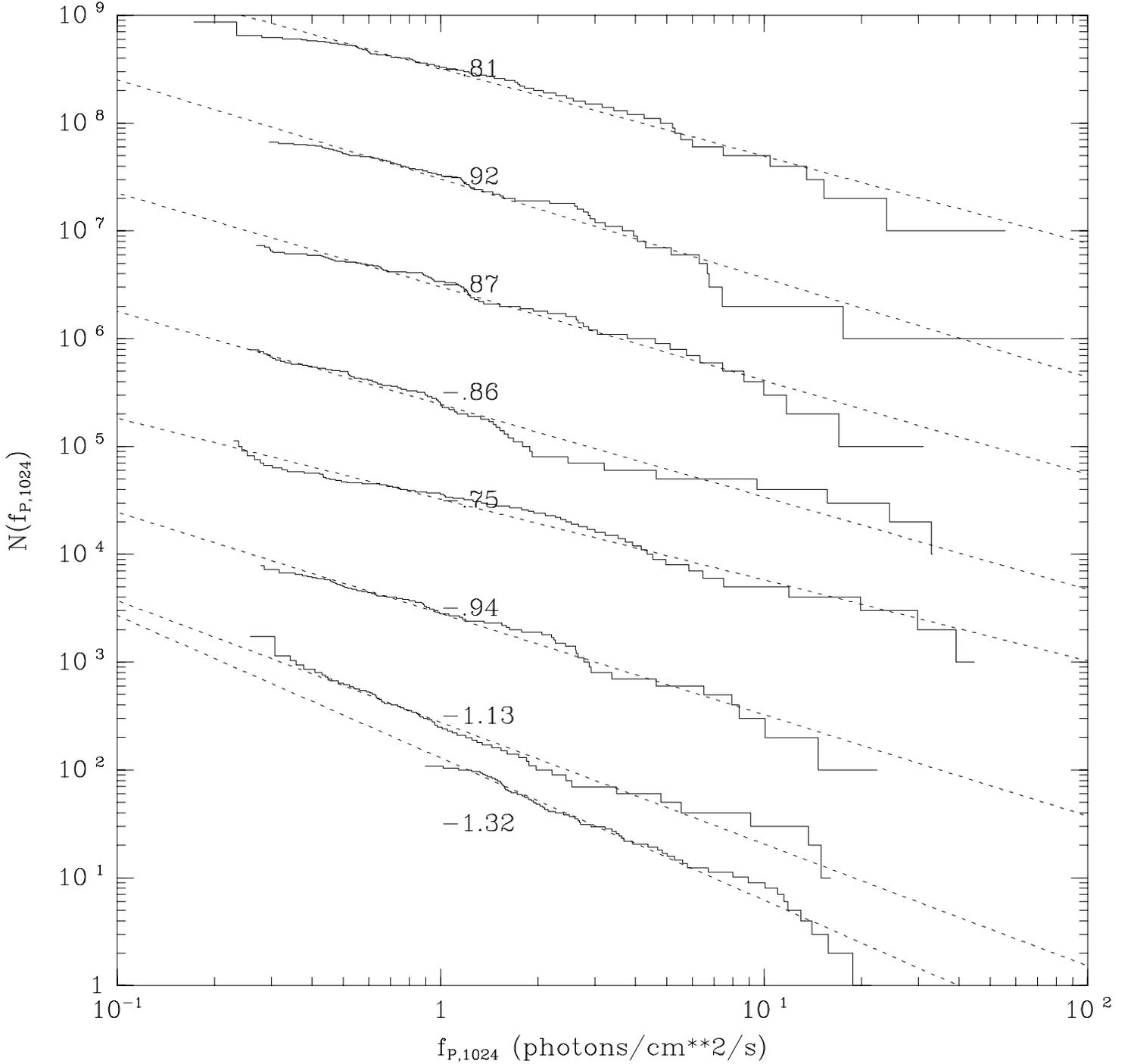

Fig. 9.— Individual cumulative distributions of $f_{P,1024}$, in 8 groups corresponding to different ranges of $T_{90}$, labeled by their logarithmic slopes. From bottom to top, the histograms correspond to groups of 65 bursts (12.5% of the total) in bins of increasing $T_{90}$. The numbers represent the approximate logarithmic slopes of the distributions. For the sake of clarity, each histogram has been scaled by a factor between 1 and $10^7$.